\documentclass[a4paper,amsmath,amssymb,10pt,aps,pre,twocolumn,floats,floatfix,superscriptaddress,showpacs]{revtex4}
\usepackage{amssymb,epsf}
\usepackage{graphicx}
\usepackage{dcolumn}
\usepackage{bm}
\usepackage{epsfig}
\usepackage{color}

\begin{document}

\title{The process of coevolutionary competitive exclusion: speciation, multifractality and power-laws in correlation}
\author{Chen-Ping Zhu}
\affiliation{College of Science, Nanjing University of Aeronautics
and Astronautics, Nanjing, 210016, P. R. China}
\affiliation{Department of Physics and Center for Computational
Science and Engineering, National University of Singapore,
Singapore, 117542, Singapore}
\author{Tao Zhou}
\affiliation{Department of Modern Physics and Institute of
Theoretical Physics, University of Science and Technology of China,
Hefei, 230026, P. R. China}
\author{Hui-Jie Yang}
\affiliation{Department of Modern Physics and Institute of
Theoretical Physics, University of Science and Technology of China,
Hefei, 230026, P. R. China}
\author{Shi-Jie Xiong}\email{sjxiong@nju.edu.cn}
\affiliation{Department of Physics and National Laboratory of
Microstructures, Nanjing University,  Nanjing,  210093,  China }
\author{Zhi-Ming Gu}
\affiliation{College of Science, Nanjing University of Aeronautics
and Astronautics, Nanjing, 210016, P. R. China}
\author{Da-Ning Shi}
\affiliation{College of Science, Nanjing University of Aeronautics
and Astronautics, Nanjing, 210016, P. R. China}
\author{ Da-Ren He}
\affiliation{College of Physical Science and Technology, Yangzhou
University, Yangzhou, 225001, P. R. China}
\author{ Bing-Hong Wang}
\affiliation{Department of Modern Physics and Institute of
Theoretical Physics, University of Science and Technology of China,
Hefei, 230026, P. R. China}
\date{\today}

\begin{abstract}
Competitive exclusion, a key principle of ecology, can be
generalized to understand many other complex systems. Individuals
under surviving pressure tend to be different from others, and
correlations among them change correspondingly to the updating of
their states. We show with numerical simulation that these aptitudes
can contribute to group formation or speciation in social fields.
Moreover, they can lead to power-law topological correlations of
complex networks. By coupling updating states of nodes with
variation of connections in a network, structural properties with
power-laws and functions like multifractality, spontaneous ranking
and evolutionary branching of node states can emerge out
simultaneously from the present self-organized model of
coevolutionary process.
\end{abstract}
\maketitle

Process of competitive exclusion\cite{Chapman} occurs in some real
systems--evolutionary branching of species in ecosystems, citations
in scientific research and designation of consumer goods are
examples among many others. It is actually a fundamental ingredient
governing main property of dynamical behaviors of systems which are
often described with complex networks\cite{Albert1} nowadays.
However, the contribution of competitive exclusion to the
interactional structure of networks and to their functional features
is not widely realized up till now.

In modeling a system, individuals are represented as nodes and
correlations among them are represented as edges of a graph.
Scale-free property\cite{Barabasi}, characterized by power-law
degree distribution, has attracted extensive attention since it
reflects a general feature of diverse systems such as the Internet,
citation networks, protein-protein interaction, and so
on\cite{Albert1}. In most previous models, dynamics of networks and
dynamics on the networks are separated. The interplay between the
formation of topological structure and functions emerge from the
network is usually neglected, which is reasonable when the structure
is independent of the dynamical states of nodes, or when these two
sides vary in rather different speeds. However, in many practical
phenomena like academic and art creation, financial transactions,
global climate fluctuation and synaptic plasticity of neuron network
in the brain\cite{Song}, both the structure and functions emerge
from the identical process, and time-dependent variations of both
individual states and local connections of nodes feed back with each
other. Therefore, novel models with coevolution
mechanisms\cite{Barat} underlying them appeared to fit for the
necessary. Unfortunately, scarcely could one produce both scale-free
structure and collective dynamics of nodes simultaneously. On the
other side, new nodes are often assumed to know the global
information of whole the growing network, which is usually
impossible for huge-size systems. In this sense it is needed to set
up models based on local interactions  to see if structure and
functions at system level will emerge from self-organized
dynamics\cite{Albert2}.

 As it is well known, competitive
 exclusion plays key role in the formation of species. There is
 strong competition among species occupying the same or nearest
 loci, surviving pressure force them drift their traits away from
 the local average level, and gradually induces evolutionary
 branching of species. Sympatric speciation\cite{branching} in an
 ecosystem is a recent
focus of naturalists. It refers to the origin of two or more species
 from a single local population.
Seceder model\cite{Soulier} based on a simple rule of local
third-order collision succeeded in mimicking such a process and
capturing its similarity to group formation in society. A network
version\cite{Gronlund} of it has been reported, giving rise to a
possible mechanism of community structure and clustering in social
networks.

In this paper the principle of competitive exclusion is generalized
outside the realm of ecology, the
 seceder model is modified to describe temporally
updated states of nodes and corresponding variation of connections
among them together. We show that generic natures of members in
diverse systems, \emph{i.e.}, to be different from others under the
pressure of competition, and coevolution between updating node
states and varying connection among nodes, can lead to simultaneous
emergences of evolutionary branching of individual traits,
spontaneous ranking and multifractality of node states and,
power-law topological structure of correlations in a system. In this
way we are able to understand scale-free phenomena and other
characteristics in various fields with a novel common mechanism.
Such self-organized coevolution models of scale-free network with
both structural and functional properties integrated are still few
to the best of our knowledge.

We set up the present model through three iteration rules. (1)
Network growth starts from a primitive complete graph with $m_0$
nodes. Each node $i$ on joining the network was assigned an initial
state with a random real number $w(i)$ uniformly distributed in
$(0,1)$. At each time step, a new node $i'$ is added to the
preexisting network. It gives out $m$ edges $(m < m_0)$ to old nodes
arbitrarily. (2) At every step, each node $ i $ counts $\bar{w}(i)$
--the average of state values $w(j)(j\ne i)$ over its nearest linked
neighbors, from them it picks up the one whose $w(j)$ makes the
maximum distance from average $\bar{w}(i)$, i.e. $J_{max}(i)$
corresponds to $max|{w(j)-\bar{w}(i)}|$, then, a randomly selected
node $j$ among the nearest neighbors of $i$ is chosen as the
offspring of $J_{max}(i)$, called $J_{sed}(i)$. Different from
original seceder model\cite{Soulier}, it is kept at its own site
and, with its state variable updated as
$w(J_{sed}(i))=w(J_{max}(i))+\delta$, where random number $\delta
\in (0,1) $ is also uniformly distributed and with positive
numerical range for wider applications. Obviously $w(i)$ here can be
accounted as a time-dependent non-decreasing
fitness\cite{Dorogovtsev1}. (3) For the new comer node $i'$ at every
step, together with its 'young' enough fellows (i.e. $i'-i\le \Delta
I$, with $\Delta I$ a given integer constant implicating aging
effect\cite{Dorogovtsev2}, hereafter we call them $I$ altogether for
convenience). Search seceders for all I's neighbors j. When
$w(J_{sed}(j))/w(I) \ge h$, where $h$ is a given value of threshold,
a new edge is added between node $J_{sed}(j)$ and $I$(Double links
and self-loops are forbidden). Meanwhile, an edge linking such node
$I$ and its neighbor $j$ is removed if the condition $w(j)/w(I)<h$
or $w(I)/w(j)<h$ is satisfied. Finally, if any node $i$ becomes
isolated due to edge-cutting, directly link it to its seceder
$J_{sed}(i)$. The threshold description of correlation adopted here
is widely used in modeling complex systems\cite{Tsonis}.

\begin{center}
\begin{figure}
\scalebox{0.8}[0.65]{\includegraphics{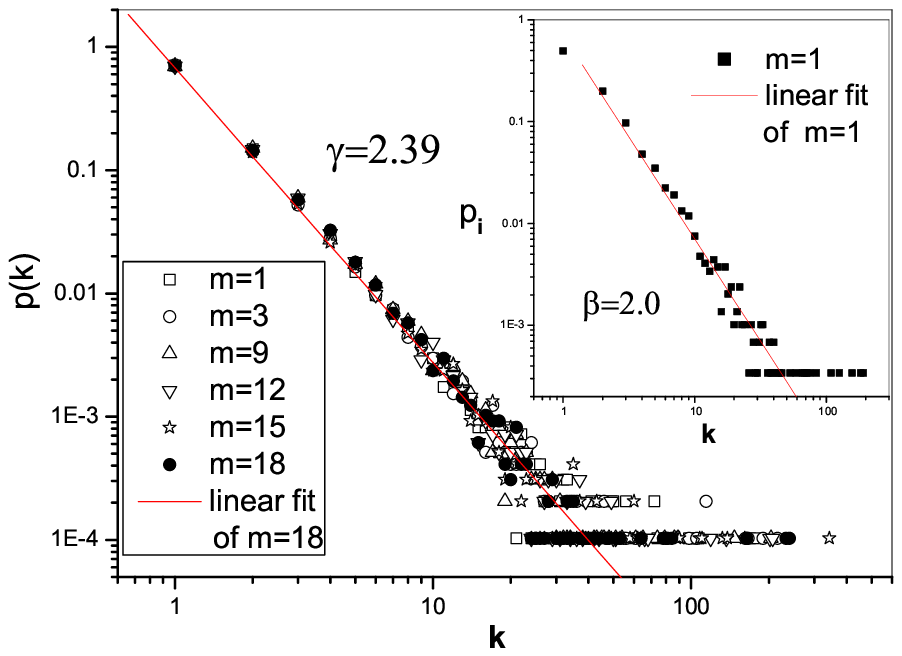}}
\scalebox{0.8}[0.65]{\includegraphics{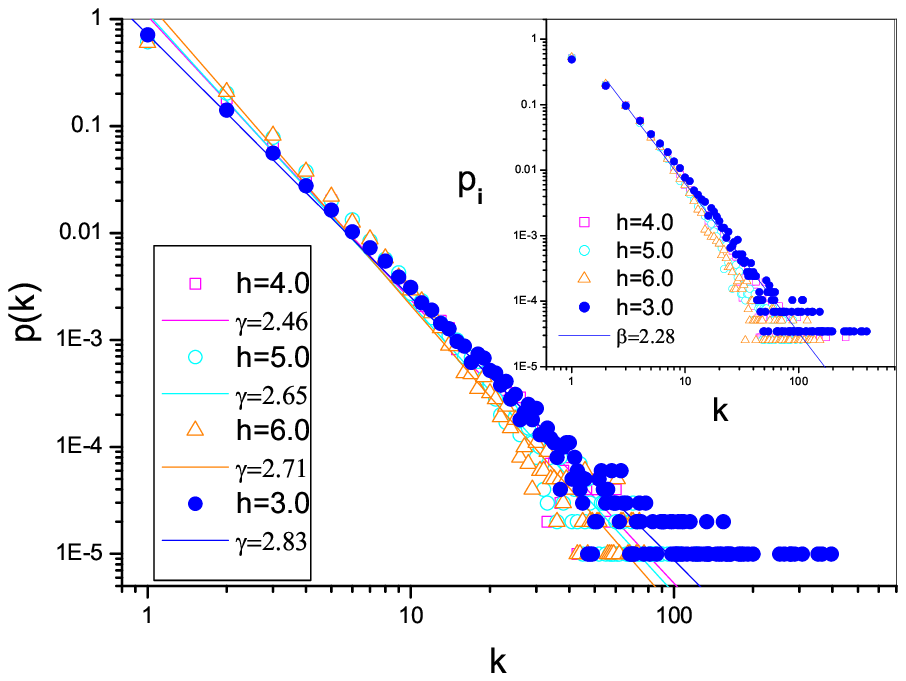}} \caption{(color
online) Power-law degree distributions of coevolving scale-free
networks. (a) Degree distributions with $h=3.0$, lines for various
values of $m$ collapse onto a single one with $\gamma=2.39$; Inset:
In-degree distribution with $\beta=2.0$ for $h=3.0$ and $m=1$.
(b)Threshold-dependent degree distributions with $h=3.0,4.0,5.0$ and
$6.0$ respectively. Inset: Corresponding in-degree distributions.
$N=10^{4}$, $\Delta I=10$ and $m_0=20$ for all lines. }
\end{figure}
\end{center}

Actually the iteration rules of the model are abstracted from
observation to real systems. In art creation and scientific
research, people have generic tendency to create new works so that
they behave differently from others. Sparkles from collision of
opinions with large difference often result in creation. As well
known, scholars are often under the pressure of publication. Papers
with the same or very similar viewpoint, method and results to
existing ones have less chance to get published. Here we see the
competition exclusion promotes prosperity of scientific research.
Suppose a graduate student just start his academic career by joining
the research on a certain topic, usually he has to focus on some
papers after extensive searching  due to limited time, and often he
extends his reading to references of them. Generally speaking, he
needs to pay more attention to ones with sharp contrasts against his
knowledge background($w(i)$), and understand lately published
literature ($w(jsed(j))$) to inspire new ideas for his own paper.
But in the reading he may be restrained within the ability of his
understanding. Therefore, it is natural to predict a suitable range
of threshold ratios within which papers with state values
$w(Jsed(j))$ would be cited (connected). And papers in selective
reading based on one's local sight are likely to be cited, forming
increased in-degree of those ones. On the opposite, papers(on the
node state $w(j)$ ) have small difference (too low ratio of
$w(j)/w(i)$) with $w(i)$ are less cited(the link between node $i$
and $j$ is trimmed). Anyway, a recently updated node
state($w(jsed(i))$) would be more attractive to a failure(an
isolated node). Artists update themselves by continuous
creation,therefor the co-occurrence network of musicians serves
another example of competitive exclusion. We know that musicians
with similar genre are competitors for performance. Managers usually
do not intend to arrange opportunity for them to appear on the same
stage since audience prefer performance with diversity. It is
assumed that whoever created a playlist was using a certain
criterion to group artists in them. One does not normally find
concerts with a mixture of heavy rock, jazz and piano sonata,
therefore a range of thresholds is used to balance the homogeneity
and heterogeneity. As the results of coevolution, both citation
network\cite{Vazquez,Braha,Myers} and musician network\cite{Cano}
display the topology of scale-free structure although most foodwebs
do not\cite{Williams}. Suppose a man faces to job crisis, he has to
refresh himself to become non-trivial for going out of dilemma. And
he may attempt to learn from, even coalesce to a succeeded person by
recommendation of a common friend. But whether they can sustain a
close relation, it depends on whether they are mutually needed and
compensate in a proper measure (e.g. $w(i)$). In all these cases
states of nodes keep varying with time and correlations among them
change corresponding to such variations along an optimal gradient.

\begin{center}
\begin{figure}
\scalebox{0.42}[0.6]{\includegraphics{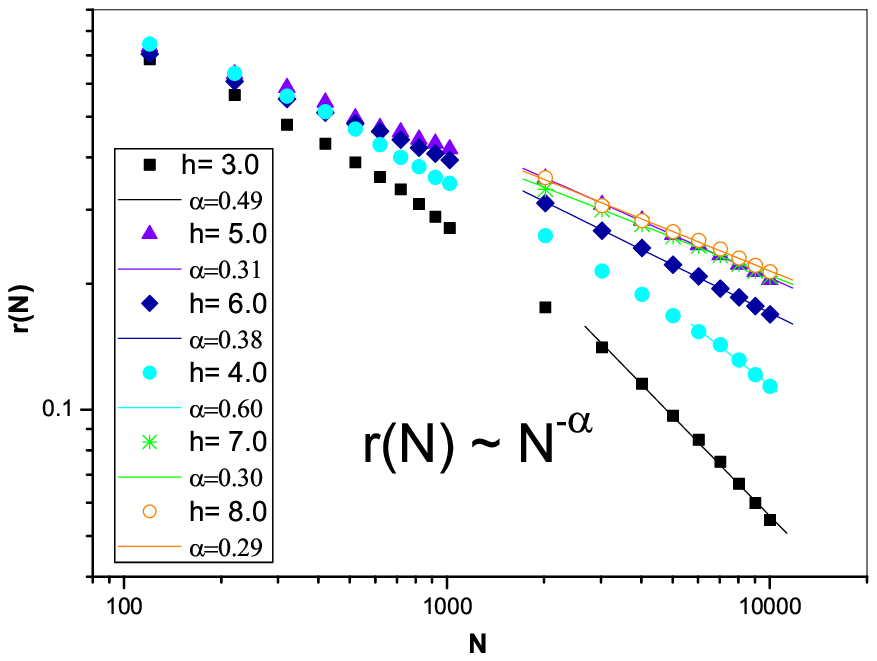}}
\scalebox{0.42}[0.6]{\includegraphics{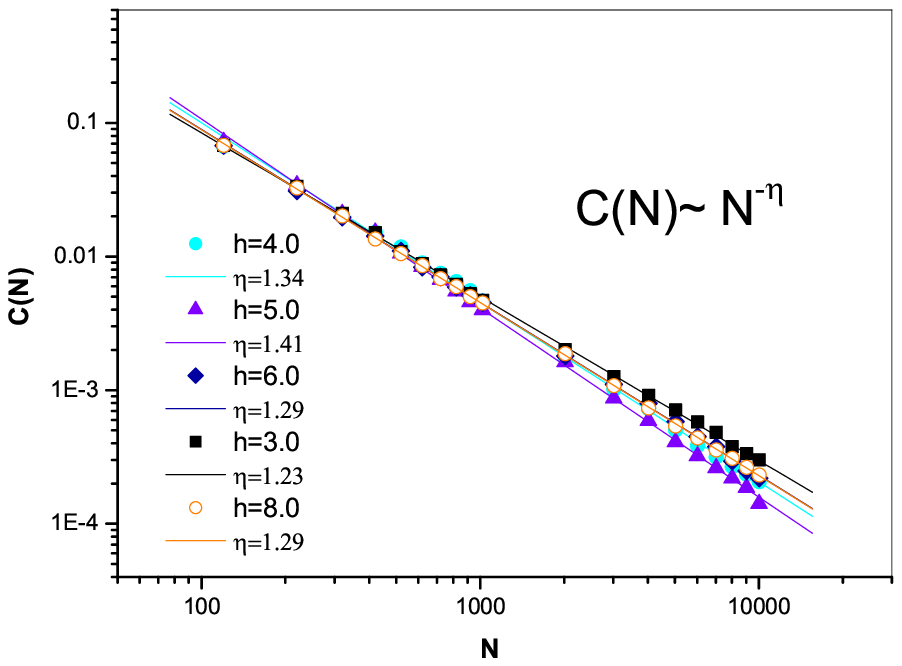}} \caption{(color
online) (a)Asymptotic size-dependent power-law decay of Pearson
coefficients: $r(N)\sim N^{-\alpha}$ for $N\agt 10^{3}$, where
$\alpha$ depends on thresholds $h$. (b)Size-dependent power-law
decay of clustering coefficient, $C(N)\sim N^{-\eta}$. Averaged on
10 realizations of network configurations.}
\end{figure}
\end{center}

 Coevolution of node states and topological connection yields most
 structural properties of complex networks by self-organization.
 Numerical simulation reveals out power-law distribution of node degree:
$p(k)\sim k^{-\gamma}$, which is illustrated in Fig. 1 a. Without
ensemble average on network configurations, it is shown that in the
case of
 $h=3.0$
 the distribution keep invariant for all values of $m$, with the slope $\gamma=2.39$.
 In-degree is counted by a node to its accepted edges from younger ones. The distribution
 also shows essential power-law as shown in the inset of Fig.1a. The slope
 of the double-logarithmic line $p_i(k)\sim k^{-\beta}$ is around $\beta=2.0$,
 which is in accordance with
 numerical results of another model\cite{Vazquez} and empirical studies\cite{Braha,Myers}.
 In Fig.1b we show the variation
 of power exponents $\gamma$ depending on correlation thresholds $h$. They lay in the range of
 (2.0, 3.0), which fits well to real complex systems. And the inset of it displays
 that essential power-law behavior of in-degree distributions also exist for different
  thresholds. The calculated
 Pearson coefficients $r$\cite{Newman} which describe degree-degree correlation of the
 network are shown in Fig.2a. They are positive reflecting statistical feature
 of social networks. Moreover, they also behave asymptotic power-law decay in the size
 of the system, i.e. $r(N)\sim N^{-\alpha}$, which is, to our knowledge, a specific feature and first predicted
 by the present model. It is expected to be verified by empirical data from real
 complex systems. Fig.2b displays size-dependent decay of clustering
 coefficients\cite{Albert1}: $C(N)\sim N^{-\eta}$.
 The exponents are $(1.2-1.4)$ corresponding to thresholds in the range of $[3.0, 8.0]$
 while for random graphs we have $\eta=1$ for comparison.

Simulations with low threshold values (\emph{e.g.}, $h=2.0$ and
$1.5$) reveal some different behaviors of the coevolution, where
iteration rules no longer lead to power-law degree distributions.
(see full and dash lines in Fig.3). However, when we allow a small
portion (ten to twenty percent) of cut-off operations not to carry
rule 3, scale-free properties can be retrieved promptly (see Fig.3).
Moreover, degree-degree correlations restore assortativity
corresponding to it. This implies that randomness may play an
essential role in the origin of scale-free behaviors since there
should be more or less relaxation on deterministic rules in complex
systems\cite{Fang}.

\begin{center}
\begin{figure}
\scalebox{0.7}[0.5]{\includegraphics{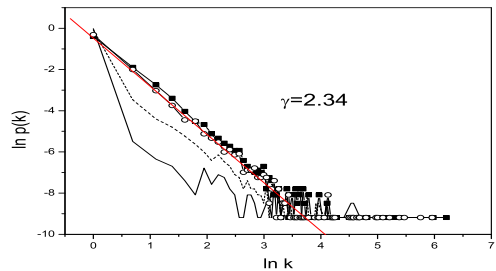}} \caption{(color
online)For low values£¨i.e.$h=2.0$, dashed line, and $h=1.5$, full
line, respectively), degree distributions deviate from power-law,
but they are retrieved by introducing ten percent(circles for
$h=2.0$) and twenty percent(squares for $h=1.5$) relaxatio
respectively on carrying out rule (3). Other parameters are the same
as those in Fig.1a.}
\end{figure}
\end{center}

\begin{figure}
 \scalebox{0.8}[0.6]{\includegraphics{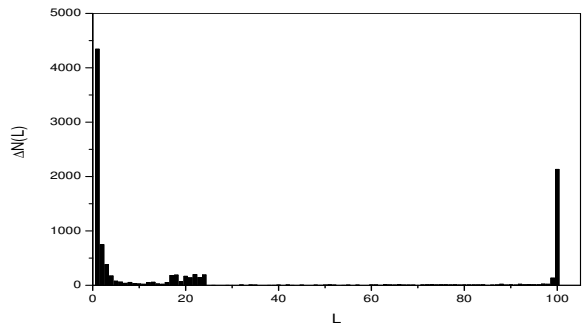}} \caption {Histogram
for numbers of nodes $¡÷N(L)$ with discrete ranks $L$ of node state
values $w(i), h=3.0$ and $m=18$, other parameters are the same as in
Fig.1a. }
\end{figure}

Ranking behavior of node states also emerge spontaneously from
coevolution. Whole the range of node states is divided into $100$
intervals in Fig.4 to show that the values are distributed quite
discontinuously. This is drastically different from uniform initial
distribution and, is comparable to group formation in original
seceder model(see Fig.1 of ref.\cite{Soulier}). Inherited from
seceder model, two prominent traits(see Fig.4) at both ends can be
regarded as the result of evolutionary branching\cite{branching}
with the tendency of elimination for mediate genotypes. Here,
species in sympatry seem to likely drift their traits away from
local average level since the strongest competition exists between
similar genotypes\cite{speciation}. Anyway, to make a scrutiny into
applicability of co-evolutionary mechanism to sympatric speciation
would be valuable. Applied to citation networks, it means that the
long term coevolution gradually eliminate the publishing chance of a
paper at middle level, instead, the population of quality tends to
be divided and shift approaching both ends. Beyond seceder
model\cite{Soulier,Gronlund}, our numerical results also give
support to the assumption of the ranking model\cite{Fortunato} of
SFN with self-organization mechanism. It is noticeable, scale-free
property as a result of coevolution can be obtained without the
prerequisite of preferential attachment on the power-law function of
prestige ranks of nodes.

\begin{center}
\begin{figure}
\scalebox{0.8}[0.65]{\includegraphics{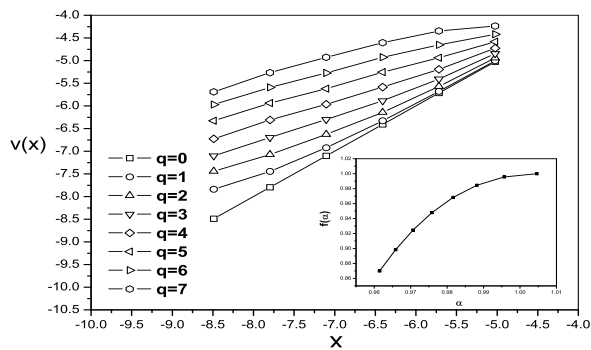}} \caption{Check the
existence of state multifractality of node according to consecutive
joining the network with box-counting method. Inset: Singularity
spectrum $f(\alpha)$ of multifractal of node state $w(i)$ for the
network with $h=3.0, m_0=20$ and $m=18$. }
\end{figure}
\end{center}

The updating process of node states induced by competitive exclusion
coupled with topological variation results in collective behavior of
nodes, which reflects characteristics of functional aspects apart
from structural ones of the network. Based on simulated data of node
states $w(i)$ which are put in order of the time sequence as node's
participation in the network, we calculate function
$V(q,d)=\sum_{l}\mu_{l}(q,d)$ln$\mu_{l}(q,d)$ with standard
box-counting technique\cite{Chahabra} for different moment $q$
versus $x=$ln $d$, where $d$ represents scales of boxes, and
$\mu_{l}$ is normalized measure of the summation over states in box
$l$. Essential linearity can be seen for at least 4-5 center lines
in Fig.5 so that the verified singularity spectrum $f(\alpha)$ of
multifractal is shown in its inset. Interestingly, the present work
gives another example of long-range correlated gradient-driven
growth of a multifractal entity\cite{Makse} with scale-free network
as its inherent skeleton. The multifractality of node states is
found to emerge accompanied with scale-free property of the
structure and vanishes correspondingly. We have verified the
correspondence between two properties in the range of $m_0 \in
[15,50]$, and $\Delta I\in [5,15]$. Therefore, the present model
suggests a common mechanism of scale-free structure of social
systems together with their multifractality and assortativity as
well.

Simultaneous emergences of macroscopic properties on both structural
and functional sides also enable us to understand functions in
coordination with the Internet, world wide spatial distribution of
population\cite{Makse} with all kinds of transport and communication
networks connecting resident sites being complex networks among
which some ones are coevolution SFNs, middle latitude climate
network\cite{Tsonis}, citation network\cite{Vazquez,Braha,Myers},
number distribution of species in ecological networks\cite{Iudin},
musician networks\cite{Cano}, and diversity maintenance method for
evolutionary optimization algorithms\cite{diversity,Myers}, on a
novel platform of coevolution with alterable details. Actually, it
relies on the mechanism with another type of preferential attachment
of node state correlation but does not explicitly depend on node
degree\cite{Barabasi,Vazquez,DSouza}, which distinguishes itself
from previous ones. Starting from but outreaching seceder model, we
can account generic natures of individuals--to update states to
self-adapt the competitive exclusion, and correlations among them
change correspondingly--as driving force in self-organization of
some evolutionary complex systems characterized by power-law
distributions of various topological quantities and specific
functions.

We acknowledge partial support from the National Science Foundation
of China (NSFC) under the Grant Nos. 70471084, 10474033, 10635040
and 60676056. CPZ  and BHW thank the hospitable accommodations of
Bao-Wen Li in NUS, and suggestion from Bing Li. SJX and BHW
acknowledge the support by National Basic Science Program of China
Project Nos. 2005CB623605, 2006CB921803 and 2006CB705500. DNS
acknowledges Foundation of NCET04-0510.

{}

\end{document}